\pdfoutput=1
\documentclass[aps,twocolumn,pra,tightenlines,floatfix,showpacs,superscriptaddress]{revtex4-1}

\usepackage{graphicx}
\usepackage{epstopdf}
\usepackage[english]{babel}
\usepackage{amsmath}
\usepackage{amssymb}
\usepackage{mathtools}
\usepackage{times}
\usepackage{appendix}
\usepackage{bm}
\usepackage{float}
\usepackage{color}
\usepackage{longtable}
\usepackage{url}
\usepackage[usenames,dvipsnames]{xcolor}
\usepackage[colorlinks=true,linkcolor=Blue,urlcolor=Blue,citecolor=Blue]{hyperref}

\graphicspath{{./Figures}}

\begin{document}

\title{Phase-space geometric Sagnac interferometer for rotation sensing}

\author{Yanming Che}
\affiliation{Zhejiang Institute of Modern Physics and Department of Physics,
Zhejiang University, Hangzhou, Zhejiang 310027, China}

\author{Fei Yao}
\affiliation{Zhejiang Institute of Modern Physics and Department of Physics,
Zhejiang University, Hangzhou, Zhejiang 310027, China}

\author{Hongbin Liang}
\affiliation{Zhejiang Institute of Modern Physics and Department of Physics,
Zhejiang University, Hangzhou, Zhejiang 310027, China}

\author{Guolong Li}
\affiliation{Zhejiang Institute of Modern Physics and Department of Physics,
Zhejiang University, Hangzhou, Zhejiang 310027, China}

\author{Xiaoguang Wang}
\email{xgwang1208@zju.edu.cn}
\affiliation{Zhejiang Institute of Modern Physics and Department of Physics,
Zhejiang University, Hangzhou, Zhejiang 310027, China}
\affiliation{Graduate School of China Academy of Engineering Physics, Beijing, 100193, China}

\date{\today}

\begin{abstract}
Quantum information processing with geometric features of quantum states may provide promising
noise-resilient schemes for quantum metrology. In this work, we theoretically explore phase-space
geometric Sagnac interferometers with trapped atomic clocks for rotation sensing, which could be
intrinsically robust to certain decoherence noises and reach high precision. With the wave guide
provided by sweeping ring-traps, we give criteria under which the well-known Sagnac phase is a
pure or unconventional geometric phase with respect to the phase space. Furthermore, corresponding
schemes for geometric Sagnac interferometers with designed sweeping angular velocity and
interrogation time are presented, and the experimental feasibility is also discussed. Such geometric
Sagnac interferometers are capable of saturating the ultimate precision limit given by the quantum
Cram\'er-Rao bound.
\end{abstract}

\maketitle

\section{Introduction}

\vspace*{-1.5ex}
Coherent manipulation of atomic clock states can be used to sense rotation
of a reference frame~\cite{DegenRMP}. By enclosing a finite area with the two distinct internal states in
real space, a Sagnac phase gate is constructed, which encodes the rotation frequency into the qubit
phase as a matter-wave Sagnac phase. With quantum resources like coherence and entanglement, such
quantum Sagnac interferometers are expected to achieve higher precision and sensitivity\cite{DegenRMP}.
However, open system effects, e.g., decoherence caused by inevitable noises, may reduce the fidelity
of the Sagnac phase gate and therefore the expected sensing precision cannot be reached.
On the other hand, geometric quantum gates have been studied
theoretically~\cite{ZanardiPLA1999,AndersPRA2000,DuanScience2001,WangPRL2001,*WangXPRA2002,
MatsumotoPRL2001,SLZhuPRL2002,ZanardiPRA2003,*ZhengPRA2004,WangZPRA2009,PechalPRL2012,ZhuPRL2003}
and demonstrated in
experiments~\cite{JonesNature2000,WinelandNature2003,DuPRA2006,AbdumalikovNature2013,FengPRL2013,SilviaNC2014,HWangNC2017}
for quantum computation. Compared to the dynamic phase, the geometric phase only depends on global
geometric features (e.g., area, volume, genus, etc.) of the state manipulation in the phase space.
Consequently, it is intrinsically immune to local noise perturbations which preserve these
geometric features~\cite{CarolloPRL2003}, and provides a promising paradigm to construct various
high-fidelity quantum phase gates. Therefore, it would be motivating to attempt to harness such 
geometric properties for high-precision quantum sensing.

Stevenson \emph{et al.} proposed a pioneering scheme for quantum rotation sensing with trap-guided
atomic clock states in Ref.~\cite{StevensonPRL2015}, and similar schemes were later considered in
Ref.~\cite{HainePRL2016} with Fisher information analyses and in Ref.~\cite{HelmPRL2018} with
spin-orbital coupling. However, the nature of the Sagnac phase $\phi_S$, i.e., whether the phase
shift is dynamic, geometric, or both, has not been clarified. And also, the fidelity and robustness of
such Sagnac phase encoding protocols under decoherence were not investigated either.

In this paper, we explore phase-space geometric quantum rotation sensing with
trapped atomic clocks, which could be potentially noise resilient and achieve high sensitivity.
With the wave guide provided by sweeping ring-traps as in Ref.~\cite{StevensonPRL2015}, we first
present the exact relation between the interferometer phase and the well-known Sagnac phase, which
could be significant in experiments, in particular for nonadiabatic interrogation cases. Then we
provide criteria under which the Sagnac phase is a pure or unconventional geometric phase~\cite{ZhuPRL2003}
with respect to the phase space. Corresponding schemes for such geometric Sagnac interferometers
with designed sweeping angular velocity and interrogation time are presented. The pure geometric
scheme would be easier to be realized in experiments in completely adiabatic guiding procedures,
while for nonadiabatic and intermediate regimes, the unconventional geometric counterparts could be
more accessible. Our results should be instrumental in experimentally implementing noise-resilient
geometric quantum rotation sensing with trapped atomic clocks.

This paper is organized as follows. In Sec.~\ref{sec:ModelandPhase} we briefly review the basic
interferometric scheme proposed in Ref.~\cite{StevensonPRL2015} and then establish the relationship
between the interferometer phase and the well-known Sagnac phase. In Sec.~\ref{sec:GeoSagnacPhase},
we investigate the geometric and dynamic components of the Sagnac phase in the phase space and give
criteria for pure and unconventional geometric Sagnac phases, followed by proposing
corresponding noise-resilient geometric Sagnac interferometer schemes. The experimental feasibilities
are also analyzed. In Sec.~\ref{sec:Precision}, the precision limit and sensitivity
given by the quantum Cram\'er-Rao bound are discussed. Finally, we conclude our work in Sec.~\ref{sec:Conclusion}.
\begin{figure*}[t]
\includegraphics[height=1.7in,width=5.3in,clip]{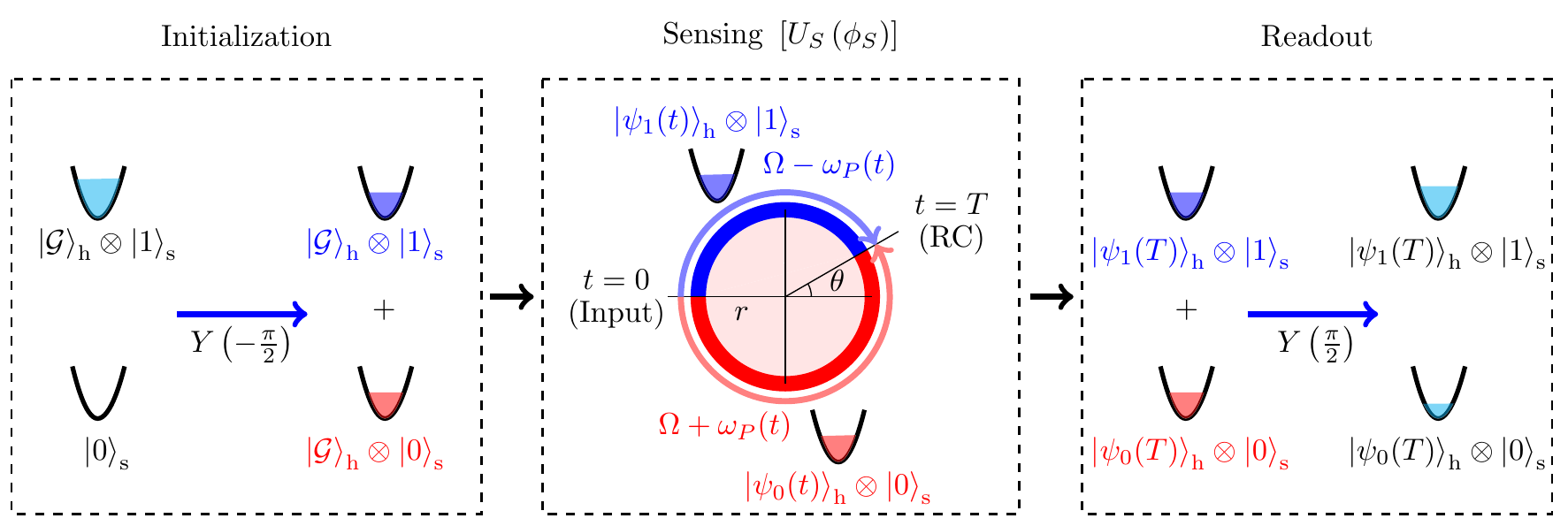}
\caption{(Color online) Schematic protocol of Sagnac interferometer with trapped atomic clock
states for rotation sensing~\cite{StevensonPRL2015}. The protocol consists of initialization,
sensing and readout, where $Y\left(\pm \pi/2\right)$ denotes the $\pi/2$ pulses and
the clock states are $\left|0\right\rangle_{\mathrm{s}}=\left|\uparrow\right\rangle$ and
$\left|1\right\rangle_{\mathrm{s}}=\left|\downarrow\right\rangle$, respectively, and
$\left|{\cal{G}}\right\rangle_{\mathrm{h}}$ is the ground state of atoms in the harmonic trap,
with the subscript $\mathrm{h}$ ($\mathrm{s}$) denoting the harmonic oscillator (spin) subspace.
In the sensing period, the atoms in two traps are coherently split at $t=0$ and are counter-transported
along a circular path of radius $r$ with respective angular velocity $\Omega \pm \omega_P(t)$ in
the inertial frame ${\cal{K}}$, and are recombined (RC) at time $T$. By properly designing the
$\omega_P(t)$ profile and the interrogation time $T$, a Sagnac phase gate $U_S\left(\phi_S\right)$
can be obtained, where $\phi_S=2\pi m r^2 \Omega/\hbar$ is the Sagnac phase. The rotation frequency
$\Omega$ can be read out from the population information after applying another $\pi/2$ pulse in the
readout stage.}
\label{fig:SagnacProtocol}
\end{figure*}

\vspace*{-1.5ex}
\section{Model and Interferometer Phase}
\vspace*{-1.5ex}
\label{sec:ModelandPhase}
Within the basic scheme in Ref.~\cite{StevensonPRL2015},
the interferometer protocol consists of two Ramsey $\pi/2$ pulses and two identical harmonic traps which
counter-transport the clock states $\left|\uparrow\right\rangle$ and $\left|\downarrow\right\rangle$
along circular paths of radius $r$ in the $xy$ plane, with respective sweeping angular velocity
$\pm \omega_P(t){\bf{z}}$ [$\omega_P(t) \ge 0$] in the rotating frame $\cal{R}$. And $\cal{R}$
rotates in an angular velocity ${\bf{\Omega}}=\Omega{\bf{z}}$ with respect to an inertial frame
${\cal{K}}$. See Fig.~\ref{fig:SagnacProtocol} for a schematic illustration. The interrogation time
$T$, when the two components are recombined for readout, is given by $\int_0^T \omega_P(t) \mathrm{d}t = \pi$.
The unitary time-evolution operators $U_{\uparrow}(T)$ and $U_{\downarrow}(T)$ for
the two respective paths form a Sagnac phase-encoding gate $U_S\left(\phi_S\right)$ at $t=T$, which
imprints the Sagnac phase into the qubit phase.
Formally, the interferometer protocol can be written as~\cite{RefPhaseNotes}
\begin{equation}
\label{eq:Protocol}
V(T) = Y\left(\frac{\pi}{2}\right)U(T)Y\left(-\frac{\pi}{2}\right),
\end{equation}
where $Y\left(\phi\right)=\mathrm{exp}\left(-i\phi \sigma_y/2\right)$ with $\sigma_y$ being
the Pauli matrix, and $U(T)={\cal{T}} \mathrm{exp}\left[-i\int_0^T H(t)\mathrm{d}t/\hbar\right]$ with
${\cal{T}}$ being the time ordering operator and $H(t)=H_0(t)\Pi_0 + H_1(t)\Pi_1$ being the Hamiltonian,
where we have used the notation
$\left|0\right\rangle_{\mathrm{s}} \left(\left|1\right\rangle_{\mathrm{s}}\right)=\left|\uparrow\right\rangle \left(\left|\downarrow\right\rangle\right)$
and $\Pi_0 \left(\Pi_1\right)=\left|0\right\rangle_{\mathrm{ss}} \hspace{-0.6ex} \left\langle0\right|\left(\left|1\right\rangle_{\mathrm{ss}} \hspace{-0.6ex} \left\langle1\right|\right)$, with the subscript $\mathrm{s}$ denoting the (pseudo)spin subspace.
It can be shown directly that (See Appendix \ref{apped:U})
\begin{equation}
\label{eq:UExpression}
U(T)=U_0(T)\Pi_0+U_1(T)\Pi_1,
\end{equation}
where
$U_{\eta}(T)={\cal{T}} \mathrm{exp}\left[-i\int_0^T H_{\eta}(t)\mathrm{d}t/\hbar\right]$
for $\eta=0, 1$ and $H_{\eta}(t)$ is the time dependent single component Hamiltonian.

For the Sensing period in the interferometer scheme shown in Fig.~\ref{fig:SagnacProtocol}, if the
degrees of freedom in the axial and radial directions for atoms in the harmonic trap are tightly
confined, then the time evolution can be described by the one-dimensional model in Ref.~\cite{StevensonPRL2015},
where the Hamiltonian for the $\left| \eta\right \rangle_{\mathrm{s}}$ state atoms in the stationary
reference frame relative to the transporting harmonic trap is given by
\begin{equation}
\label{eq:Hamiltonian}
H_{\eta}(t) = \hbar \omega_0 \left(a^{\dagger}a+\frac{1}{2}\right) + i\lambda_{\eta}(t)\left(a-a^{\dagger}\right),
\end{equation}
where $\omega_0$ is the trap frequency and $a\left(a^{\dagger}\right)$ is the annihilation (creation)
operator for the trap mode.
The second term in Eq.~(\ref{eq:Hamiltonian}) represents the drive acting on the harmonic oscillator
induced by the rotation of the frame, where
$\lambda_{\eta}(t)=\sqrt{\frac{m\hbar\omega_0}{2}}r\left[\Omega + (1-2\eta) \omega_P (t)\right]$,
with $m$ being the particle mass and $\Omega$ the rotation frequency to be measured.
$\omega_P (t) \ge 0$ for $t \in [0, T]$ is the experimentally designed sweeping angular velocity whose
temporal profile determines the interferometer phase and the signal contrast, which
will be shown below. The sweeping angular velocity $\omega_P (t)$ can be further extended to a
function ${\cal{W}}_P(t)$ defined on the whole real-time axis $t \in \left[-\infty, +\infty\right]$,
with ${\cal{W}}_P(t) = \omega_P (t)$ for $t \in [0, T]$ and ${\cal{W}}_P(t)=0$ for the other,
and the frequency spectrum of ${\cal{W}}_P(t)$ can be obtained from its Fourier transform
\begin{equation}
\label{eq:FourierSpectrum}
\widetilde{{\cal{W}}}_P(\omega)=\frac{1}{\sqrt{2\pi}} \int_{-\infty}^{+\infty}{\cal{W}}_P(t) \mathrm{exp}\left(-i\omega t\right) \mathrm{d}t.
\end{equation}

The Hamiltonian in Eq.~(\ref{eq:Hamiltonian}) describes a driven harmonic oscillator and the
corresponding time evolution operator at time $t$ is given by (see Appendix \ref{apped:U} for
detailed derivations)
\begin{equation}
\label{eq:U}
U_{\eta}(t)=D\left[\alpha_{\eta}(t)\right] e^{-i\omega_0 a^{\dagger}a t} e^{i\left[\phi_{\eta}(t) - \omega_0 t/2\right]},
\end{equation}
where $D\left(\alpha\right)=\mathrm{exp}\left(\alpha a^{\dagger}-\alpha^* a\right)$ is the displacement
operator for the harmonic oscillator, with
\begin{eqnarray}
\label{eq:AlphaPhi}
\alpha_{\eta} (t) &=&  -\frac{1}{\hbar} \int^t_0 \lambda_{\eta}(\tau) \mathrm{exp}\left[i\omega_0 \left(\tau-t\right)\right]  \mathrm{d} \tau, \nonumber \\
\phi_{\eta}(t) &=& \frac{1}{\hbar^2}\int_0^t \int_0^{\tau_1} \lambda_{\eta}(\tau_1) \lambda_{\eta}(\tau_2) \sin \left[\omega_0 \left(\tau_1 - \tau_2 \right)\right]
\mathrm{d}\tau_2 \mathrm{d}\tau_1. \nonumber \\
\end{eqnarray}
If the initial states for both $\left|0\right\rangle_{\mathrm{s}}$ and
$\left|1\right\rangle_{\mathrm{s}}$ components in the harmonic trap are in the ground state (vacuum)
$\left|{\cal{G}}\right\rangle_{\mathrm{h}}$ which is defined by $a \left|{\cal{G}}\right\rangle_{\mathrm{h}}=0$,
where the subscript ${\mathrm{h}}$ denotes the harmonic trap subspace, then the state at time $t$ is
given by
\begin{eqnarray}
\left|\psi_{\eta}(t)\right\rangle_{\mathrm{h}} &=& U_{\eta}(t) \left|{\cal{G}}\right\rangle_{\mathrm{h}} \nonumber \\
&=& \left|\alpha_{\eta}(t)\right\rangle_{\mathrm{h}} \mathrm{exp}\left[i\left(\phi_{\eta}(t) - \omega_0 t/2\right)\right],
\end{eqnarray}
where $\left|\alpha_{\eta}(t)\right\rangle_{\mathrm{h}}$ is the coherent state which is eigenstate
of $a$ with eigenvalue $\alpha_{\eta}(t)$. See the sensing period in Fig.~\ref{fig:SagnacProtocol}.

For the initial state $\rho(0) = \left|{\cal{G}}\right\rangle_{\mathrm{hh}} \hspace{-0.65ex} \left\langle
{\cal{G}}\right| \otimes \left|1\right\rangle_{\mathrm{ss}} \hspace{-0.6ex} \left\langle1\right|$
of the interferometer, the readout state reads $\rho(T)= V(T)\rho(0)V^{\dagger}(T)$.
The reduced density matrix for the spin subspace is given by $\rho_{\mathrm{s}}(T)= \mathrm{Tr_{h}} \rho(T)$
with $\mathrm{Tr}$ being the trace operation, and reads
\begin{equation}
\label{eq:ReducedDM}
\rho_{\mathrm{s}}(T)= \frac{1}{2}\left[{\cal{I}}_2-\mathrm{Re}\left({\cal{C}}_{1, 0}\right) \sigma_z
- \mathrm{Im}\left({\cal{C}}_{1, 0}\right) \sigma_y \right],
\end{equation}
where ${\cal{I}}_2$ is the two-dimensional identity matrix, ${\cal{C}}_{j, k}={}_{\mathrm{h}}\hspace{-0.5ex} \left\langle \alpha_j(T) | \alpha_k(T) \right \rangle_{\mathrm{h}} \mathrm{exp}\left\{-i\left[\phi_j(T)-\phi_k(T)\right]\right\}$
with $j, k \in \{0, 1\}$ and $\mathrm{Re}\left(\mathrm{Im}\right)$ denotes the real (imaginary)
part. $\sigma_z = \left|0\right\rangle \left\langle 0\right| - \left|1\right\rangle \left\langle 1\right|$
and $\sigma_y = i\left(\left|1\right\rangle \left\langle 0\right| - \left|0\right\rangle \left\langle 1\right|\right)$
are the Pauli operators. Therefore, the measurement signal, e.g., the population difference, is
given by
\begin{equation}
\label{eq:Signal}
\quad_{\mathrm{s}} \hspace{-0.3ex} \left\langle \sigma_z \right \rangle _{\mathrm{s}}= -|{\cal{C}}_{1, 0}|\mathrm{cos} \left(\phi_I\right),
\end{equation}
where the modulus $|{\cal{C}}_{1, 0}|=\mathrm{exp}\left(-|\Delta \alpha|^2/2\right)$ gives the
signal contrast, with
$\Delta \alpha=\alpha_0 (T)-\alpha_1(T) \propto \widetilde{{\cal{W}}}_P^* \left(\omega_0\right)$,
and $\phi_I=\mathrm{arg}\left({\cal{C}}_{1, 0}\right)$ is the interferometer phase, with
$\mathrm{arg}$ denoting the argument. From straightforward calculation one obtains the
interferometer phase $\phi_I$, which is given by (see Appendix \ref{apped:Phase})
\begin{equation}
\label{eq:InterferometerPhase}
\phi_I= \phi_S \left\{1-\sqrt{\frac{2}{\pi}} \mathrm{Re}\left[\widetilde{{\cal{W}}}_P \left(\omega_0\right)\right]\right\},
\end{equation}
where $\phi_S=2\pi m r^2 \Omega/\hbar$ is the well-known Sagnac phase and it can be further shown
that $0 \le \phi_I \le 2\phi_S$. In contrast to Ref.~\cite{StevensonPRL2015}, the
phase of the interferometer in Eq.~(\ref{eq:Signal})
is indeed \emph{dependent} on the Fourier components of $\widetilde{{\cal{W}}}_P \left(\omega\right)$
at the trap frequency $\omega_0$ and therefore depends on the temporal profile of $\omega_P(t)$. This
result is experimentally relevant because the form of $\omega_P(t)$ profile will affect the
interference fringes. Now we arrive at the condition
\begin{equation}
\label{eq:AdiaCondition}
\mathrm{Re}\left[\widetilde{{\cal{W}}}_P \left(\omega_0\right)\right]=0,
\end{equation}
under which $\phi_I=\phi_S$, i.e., the interferometer phase is exactly the Sagnac phase.

It should be noted that Eq.~(\ref{eq:AdiaCondition}) could be satisfied when
the guiding procedure is performed in an adiabatic fashion, i.e., $T \gg 2\pi/\omega_0$.
For example, for a constant $\omega_P(t)=\pi/T$, we have
$\mathrm{Re}\left[\widetilde{{\cal{W}}}_P \left(\omega\right)\right]=\sqrt{\pi/2}
\mathrm{sin}\left(\omega T\right)/\left(\omega T\right)$. Consequently, the amplitude of the
frequency distribution decreases with $\omega$ in power law and for large $\omega_0$ located
far away from the typical spectrum width of the order $1/T$, the contribution from the real
part in Eq.~(\ref{eq:InterferometerPhase}) approaches $0$. On the other hand, for nonadiabatic
guiding procedures with $T \sim \omega_0^{-1}$ (if possible in future experiments), the sweeping
angular velocity and the interrogation time should be properly designed such that $\omega_0$
lies in the node of the frequency spectrum.

\vspace*{-1.5ex}
\section{Phase-Space Geometric Sagnac Interferometers}
\vspace*{-1.5ex}
\label{sec:GeoSagnacPhase}
Up to now we have obtained the relationship between the interferometer phase and the
Sagnac phase, but the dynamic and geometric origins of the phase in the phase space have
not been investigated yet. Next we analyze different components in the Sagnac phase
$\phi_S$ and explore unconventional geometric~\cite{ZhuPRL2003,DuPRA2006} Sagnac interferometers,
which could be potentially resilient to noises and are promising for reaching high-precision
quantum rotation sensing.

Quantum geometric phase is the phase change associated with holonomic transformation in quantum state
space~\cite{SimonPRL1983,AnandanPRD1987,AharonovPRL1987,HWangNC2017}, which was studied by
Berry~\cite{Berry1984} for adiabatic cyclic motion and by Aharonov and Anandan~\cite{AharonovPRL1987}
for any cyclic evolution. Here for the spin state $\left| \eta \right\rangle_{\mathrm{s}}$ ($\eta=0, 1$), the
total phase change for quantum evolution in each harmonic trap can be divided into dynamic and geometric
components, which are given by~\cite{ZhuPRL2003,CarolloPRL2003}
\begin{equation}
\label{eq:DPhase}
 \gamma^{\mathrm{d}}_{\eta}(T)=-\int_0^T
 \left\langle \psi_{\eta}(t) |H_{\eta}(t)| \psi_{\eta}(t) \right \rangle \mathrm{d}t,
\end{equation}
and
\begin{eqnarray}
\label{eq:GPhase}
 \gamma^{\mathrm{g}}_{\eta}(T) = \frac{i}{2}\int_{\Gamma_{\eta}} \left(\alpha_{\eta}^*
 \mathrm{d} \alpha_{\eta} - \alpha_{\eta}  \mathrm{d} \alpha^*_{\eta} \right)-
 \mathrm{arg}\left[\left\langle \alpha_{\eta}(T) |{\cal{G}} \right \rangle\right], \nonumber \\
\end{eqnarray}
respectively. Note that here in Eqs.~(\ref{eq:DPhase}) (\ref{eq:GPhase}) and hereafter we will drop the
subspace subscript $\mathrm{h}$ for convenience. From Eq.~(\ref{eq:GPhase}), one sees that the
geometric phase consists of two parts, where the first is $-2$ times the area~\cite{DoubleAreaNotes}
subtended by the path $\Gamma_{\eta}=\{\alpha_{\eta}(t)| t \in [0, T]\}$ of motion in the phase
space and the second is the argument of the overlap between the initial and final coherent states. Define
$\Delta \gamma^{\mathrm{d(g)}}=\gamma_0^{\mathrm{d(g)}}(T)-\gamma_1^{\mathrm{d(g)}}(T)$
as the dynamic (geometric) phase difference of the interferometer, which satisfies
$\phi_I = \Delta \gamma^{\mathrm{d}} + \Delta \gamma^{\mathrm{g}}+\mathrm{arg}\left[\left\langle
\alpha_{1}(T) |\alpha_{0}(T) \right\rangle\right]$.
The third component in $\phi_I$ is given by
$\mathrm{arg}\left[\left\langle \alpha_{1}(T) |\alpha_{0}(T) \right\rangle\right]
=|\alpha_{0}(T) \alpha^*_{1}(T) |\mathrm{sin}\left\{\mathrm{arg}\left[\alpha_{0}(T)\right]
- \mathrm{arg}\left[\alpha_{1}(T)\right]\right\}$,
which only depends on the respective final positions of the $\left|0\right\rangle$ and
$\left|1\right\rangle$ state atoms in the trap when they are recombined, and also has a clear
geometric meaning. Therefore, it can be absorbed into the geometric phase difference. Consequently,
the total phase difference of the Sagnac interferometer in Eq.~(\ref{eq:InterferometerPhase}) can
be decomposed into
\begin{eqnarray}
\label{eq:PhaseDecomp}
 \phi_I = \Delta \gamma^{\mathrm{d}} + \Delta \tilde{\gamma}^{\mathrm{g}},
\end{eqnarray}
where $\Delta \tilde{\gamma}^{\mathrm{g}}=\gamma_0^{\mathrm{g}}(T)-\gamma_1^{\mathrm{g}}(T)
+\mathrm{arg}\left[\left\langle \alpha_{1}(T) |\alpha_{0}(T) \right\rangle\right]$
is the purely geometric contribution related to the area and angle differences in the phase spaces,
respectively.

Next, with a theorem, we show that by properly designing the interrogation time $T$ and the
temporal profile of the sweeping angular velocity $\omega_P(t)$, the Sagnac phase can be made a
pure or unconventional geometric phase, where by the latter we mean that the geometric Sagnac phase
also involves a dynamic component~\cite{ZhuPRL2003}. We give criteria for the phase-space
geometric Sagnac phase, followed by proposed experimentally accessible schemes for geometric
Sagnac interferometers.

\emph{Theorem.} For certain proper interrogation time $T$ and temporal profiles of
$\omega_P(t) \ge 0$ with $t \in [0, T]$ which satisfy $\phi_I=\phi_S$, there exist
nonzero $\kappa \in \mathbb{R}$ such that
\begin{eqnarray}
\label{eq:DGRatio}
 \Delta \gamma^{\mathrm{d}}=\left(\kappa - 1\right) \Delta \tilde{\gamma}^{\mathrm{g}},
\end{eqnarray}
where for $\kappa=1$, the Sagnac phase $\phi_S$ is purely geometric, and for $\kappa \ne 1$,
$\phi_S$ is an unconventional geometric phase~\cite{IntPicNote}.

\emph{Proof and examples.} With the frequency spectrum $\widetilde{{\cal{W}}}_P\left(\omega\right)$
and straightforward calculations we obtain (see Appendix \ref{apped:PhaseDecomposition} for
detailed calculations)
\begin{eqnarray}
\label{eq:PhaseComponents}
 \Delta \tilde{\gamma}^{\mathrm{g}}=\sqrt{\frac{2}{\pi}}\phi_S \xi\left(\omega_0, T\right), \ \ \
 \Delta \gamma^{\mathrm{d}}=\phi_I - \Delta \tilde{\gamma}^{\mathrm{g}},
\end{eqnarray}
where $\phi_I$ is given by Eq.~(\ref{eq:InterferometerPhase}) and
$\xi\left(\omega_0, T\right)=\omega_0 \partial_{\omega}\mathrm{Re}
\left[\widetilde{{\cal{W}}}_P\left(\omega\right)\right]_{\omega=\omega_0}- \omega_0 T
\mathrm{Im}\left[\widetilde{{\cal{W}}}_P\left(\omega_0\right)\right]$.

Under the condition $\phi_I=\phi_S$ and maximization of the contrast, which indicate
$\widetilde{{\cal{W}}}_P \left(\omega_0\right)=0$, $\kappa$ in Eq.~(\ref{eq:DGRatio}) can be
expressed as
\begin{equation}
\kappa=\sqrt{\pi/2}/\xi_0\left(\omega_0, T\right),
\end{equation}
with $\xi_0\left(\omega_0, T\right)=\omega_0 \partial_{\omega}
\mathrm{Re}\left[\widetilde{{\cal{W}}}_P\left(\omega\right)\right]_{\omega=\omega_0}$.
Therefore, for properly designed $\omega_P(t)$ and $T$, if there exists nonzero $\xi_0\left(\omega_0, T\right)$,
then this theorem holds automatically. The unconventional geometric class with
$\xi_0\left(\omega_0, T\right) \ne \sqrt{\pi/2}$
should be more generic. Below we give two examples of this class with sinusoidal and cosinusoidal
temporal profiles for $\omega_P(t)$, respectively, which could be schemes for the unconventional
geometric Sagnac interferometer. A pure geometric scheme with flat $\omega_P(t)$ profile will
also be presented in the following.

\emph{(i) Unconventional geometric Sagnac phase.} Firstly we present a designed sinusoidal angular
velocity $\omega_P(t)=\pi^2 |\mathrm{sin}\left(2\pi t/T\right)|/\left(2T\right)$
for $t \in [0, T]$ with $T=2\pi/\omega_0$, which sets $\phi_I=\phi_S$ and maximizes the contrast
at the same time, i.e., $\widetilde{{\cal{W}}}_P\left(\omega_0\right)=0$ for this situation. This
sinusoidal profile results in a nontrivial solution for $\kappa$ in Eq.~(\ref{eq:DGRatio}),
$\kappa=8/\pi^2$, and the Sagnac phase $\phi_S=8 \Delta \tilde{\gamma}^{\mathrm{g}}/\pi^2$
(see Appendix \ref{apped:GeoSagncExamples} for detailed calculations)
is an unconventional geometric phase, by which we mean that the geometric $\phi_S$ also involves a
dynamic component~\cite{ZhuPRL2003}.

Secondly, we find that the cosinusoidal angular velocity
profile (sinusoidal angular profile) used in Ref.~\cite{HainePRL2016} in order to calculate the
Fisher information, may also provide a nontrivial scheme for the unconventional geometric Sagnac
interferometer, where $\omega_P(t)=\left(\pi/T\right)\left[1-\mathrm{cos}\left(2\pi t/T\right)\right]$.
By choosing the interrogation time to be $T=2M\pi/\omega_0$, where $M=2, 3, 4,...$, we have
$\widetilde{{\cal{W}}}_P\left(\omega_0\right)=0$, and the corresponding $\kappa$ is given by
$\kappa=1-M^2$. Therefore, the Sagnac phase
$\phi_S=\left(1-M^2\right)\Delta \tilde{\gamma}^{\mathrm{g}}$ (see Appendix \ref{apped:GeoSagncExamples})
is also an unconventional geometric phase. It is worth noting that the former sinusoidal profile
case can only be performed in a nonadiabatic guiding procedure due to $T \sim \omega_0^{-1}$, while
the latter cosinusoidal case is applicable to both adiabatic and nonadiabatic scenarios,
which depends on the value of $M$ taken in $T \sim M\omega_0^{-1}$.

\emph{(ii) Pure geometric Sagnac phase.} A constant angular velocity $\omega_P(t)=\pi/T$ for $t \in [0, T]$
with $T=2K \pi/\omega_0$ gives $\widetilde{{\cal{W}}}_P\left(\omega_0\right)=0$, where $K=1, 2, 3,\cdots$,
and therefore $\phi_I=\phi_S$ and the contrast is maximized
simultaneously. The solution for $\kappa$ in Eq.~(\ref{eq:DGRatio}) is $\kappa \equiv 1$ (see Appendix \ref{apped:GeoSagncExamples}).
Furthermore, in this example $\gamma_{\eta}^{\mathrm{d}}(T)=-K \pi$ for both branches with $\eta=0$
and $1$, which comes from the zero-energy contribution. So the Sagnac phase in this case is
purely geometric.
\begin{figure}
\centerline{\includegraphics[height=2.55in,width=3.5in,clip]{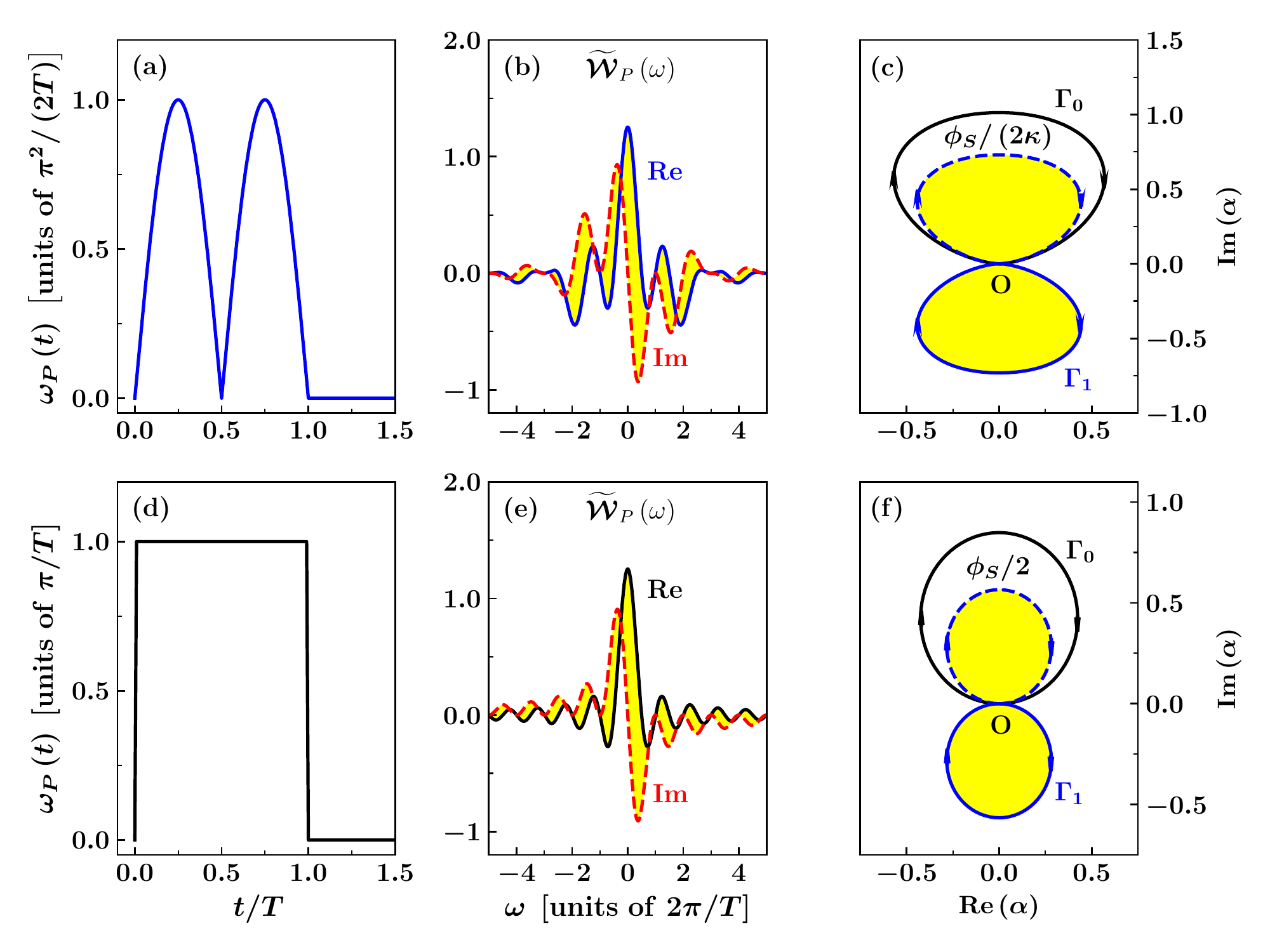}}
\caption{(Color online) Unconventional [(a)--(c)] and pure [(d)--(f)] geometric Sagnac
interferometers with atomic clock states, where (a) and (d), (b) and (e), (c) and (f) share the same
horizontal axis labels, respectively. (a) is the sinusoidal $\omega_P(t)$ profile
[in units of $\pi^2/(2T)$] in the example (i) and (d) is the flat $\omega_P(t)$ profile (in units of $\pi/T$)
in the example (ii). (b) and (e) are Fourier transform
$\widetilde{{\cal{W}}}_P\left(\omega\right)$ of (a) and (d), respectively, where $\mathrm{Re}$ ($\mathrm{Im}$)
denotes the real (imaginary) part. The horizontal frequency axes in both (b) and (e) are in units
of $2\pi/T$. (c) and (f) are the phase space paths for the atoms in the two respective traps
during the interrogation, where $\Gamma_{\eta}=\{\alpha_{\eta}(t)| t \in [0, T]\}$ is the path of
$\left| \eta \right\rangle_{\mathrm{s}}$ state atoms in the $\alpha$ plane with $\eta=0$ and $1$, respectively.
(c) is plotted with the sinusoidal $\omega_P(t)$ profile in (a), and (f) is with the flat profile
in (d). For both (c) and (f), we take $\omega_0=\hbar=m=r=1$, $\Omega=0.1\omega_0$ and $T=2\pi/\omega_0$,
and the dashed blue line denotes $-\Gamma_{1}=\{-\alpha_{1}(t)| t \in [0, T]\}$ which encloses the
same area as $\Gamma_{1}$. The area of the unfilled region inside $\Gamma_{0}$ is proportional to
$\phi_S/2$ in (c) with $\kappa=8/\pi^2$, while it is identical with $\phi_S/2$ in (f).}
\label{fig:GeoPhase}
\end{figure}

The physical pictures for above schemes are as follows: When the two branches are recombined at $t=T$,
the atoms in each trap accomplish integer numbers of cyclic evolutions, and return 
to the initial vacuum state, during which the Sagnac phase is given by the area
difference subtended by the two trajectories in the phase spaces. 
Figures ~\ref{fig:GeoPhase}(a)--\ref{fig:GeoPhase}(c) plot the unconventional geometric Sagnac phase with the
sinusoidal $\omega_P(t)$ profile and Figs.~\ref{fig:GeoPhase}(d)--\ref{fig:GeoPhase}(f) show the pure geometric
counterpart with the flat $\omega_P(t)$ profile. In Figs.~\ref{fig:GeoPhase}(b) and \ref{fig:GeoPhase}(e), to
satisfy $\phi_I=\phi_S$ and to maximize the contrast at the same time, the trap frequency $\omega_0$
is given by the positive simultaneous zeros of real ($\mathrm{Re}$) and imaginary ($\mathrm{Im}$)
parts of $\widetilde{{\cal{W}}}_P\left(\omega\right)$, which is $\omega_0 T=2\left(2L+1\right)\pi$
for Fig.~\ref{fig:GeoPhase} (b) with $L=0, 1, 2,\cdots$, and $\omega_0 T=2 K \pi$ for
Fig.~\ref{fig:GeoPhase}(e) with $K$ being a positive integer (see Appendix \ref{apped:GeoSagncExamples}).
Shown in Figs.~\ref{fig:GeoPhase}(c) and \ref{fig:GeoPhase}(f) are
the phase space paths $\Gamma_{\eta}$ for $\left| \eta \right\rangle_{\mathrm{s}}$ state atoms
during the interrogation, with $\eta=0$ and $1$, respectively. Fig.~\ref{fig:GeoPhase}(c) is
plotted with the sinusoidal $\omega_P(t)$ profile and $L=0$, and Fig.~\ref{fig:GeoPhase}(f) is with
the flat profile and $K=1$. Note that in Fig.~\ref{fig:GeoPhase}(b), only the $L=0$ case can give
a nontrivial solution for $\kappa$ in Eq.~(\ref{eq:DGRatio}). The dashed blue line
denotes $-\Gamma_{1}=\{-\alpha_{1}(t)| t \in [0, T]\}$
which encloses the same area as $\Gamma_{1}$. Therefore, the area of the unfilled region inside
$\Gamma_{0}$, which is equal to $\Delta \tilde{\gamma}^{\mathrm{g}}/2$, is identical with
$\phi_S/\left(2\kappa\right)$ with $\kappa=8/\pi^2$ in Fig.~\ref{fig:GeoPhase}(c) while it equals
$\phi_S/2$ in Fig.~\ref{fig:GeoPhase}(f), which are signatures of unconventional and pure geometric
Sagnac phases, respectively.

So we have provided criteria for geometric Sagnac interferometers in the phase space and
proposed corresponding schemes for geometric quantum rotation sensing with trap-guided atomic
clocks. It should be noted that in the completely adiabatic regime with $T \gg \omega_0^{-1}$,
the unconventional geometric phase component given by the cosinusoidal $\omega_P(t)$ in example (i) is
$\Delta \tilde{\gamma}^{\mathrm{g}} \propto \kappa^{-1} \approx 0$, and $\phi_S$ becomes nearly
completely dynamic. In this regime, the pure geometric scheme with flat $\omega_P(t) \propto T^{-1}$
in example (ii) is more accessible in experiments due to the fact that $\kappa \equiv 1$ for this scheme.
In nonadiabatic and intermediate regimes, the unconventional geometric schemes in example (i) could be
more accessible due to continuous $\omega_P(t)$ with $\omega_P(0)=\omega_P(T)=0$.

\vspace*{-1.5ex}
\section{Precision and Sensitivity}
\vspace*{-1.5ex}
\label{sec:Precision}
Now we discuss the precision and sensitivity of the above schemes. The sensitivity of
the Sagnac interferometer is limited by the uncertainty of the unbiased estimated value of
$\Omega$, which is given by
$\delta \Omega=\delta P\left(\phi_I\right) / \left|\partial_{\Omega} P\left(\phi_I\right)\right|$,
where $P\left(\phi_I\right)= _{\mathrm{s}} \hspace{-1ex} \left\langle \sigma_z \right \rangle _{\mathrm{s}}$
is the signal given in Eq.~(\ref{eq:Signal}). Straightforward calculation leads to
\begin{equation}
\label{eq:uncertainty}
\frac{1}{\left(\delta \Omega\right)^2}=\frac{\left(\frac{\partial \phi_I}{\partial \Omega}\right)^2 \mathrm{sin}^2 \phi_I}
{\left|{\cal{C}}_{1, 0}\right|^{-2}-1+\mathrm{sin}^2 \phi_I}  \le {\cal{F}},
\end{equation}
where we have $\left|{\cal{C}}_{1, 0}\right| \le 1$ and ${\cal{F}}$ is the quantum Fisher
information (QFI) which determines the ultimate precision limit for quantum sensing via the
quantum Cram\'er-Rao bound (QCRB)~\cite{Helstrom,Holevo}. For the interferometer protocol $V(T)$
and the initial state $\rho(0)$ in this paper, if the interrogation time is integer times the
trap period, i.e., $T=2K \pi/\omega_0 $ with $K=1,2,3,\cdots$, then the QFI in Eq.~(\ref{eq:uncertainty})
is given by ${\cal{F}}=\left(\partial_{\Omega} \phi_I\right)^2$~~\cite{YaoUnpublish2018}.
Therefore, the conditions for attaining the equality in Eq.~(\ref{eq:uncertainty}) and saturating
the QCRB are $\left|{\cal{C}}_{1, 0}\right|=1$ [or $\widetilde{{\cal{W}}}_P \left(\omega_0\right)=0$]
and $\omega_0 T=2K \pi$. So all the schemes proposed in the examples (i) and (ii) with $P\left(\phi_I\right)$
measurements satisfy these conditions and thus saturate the QCRB.

\vspace*{-1.5ex}
\section{Conclusion}
\vspace*{-1.5ex}
\label{sec:Conclusion}
In summary, we have proposed schemes for phase-space geometric Sagnac
interferometers with trap-guided atomic clocks, which could be potentially noise resilient
and promising for high-sensitivity rotation sensors. The pure geometric scheme is applicable
to adiabatic guiding procedures while the unconventional geometric schemes could be more accessible
in nonadiabatic situations. In addition, the established relationship between the interferometer
phase and the Sagnac phase may provide a theoretical basis of evaluating the scale factor for the
Sagnac interferometer, which is crucial for the accuracy of atomic sensors.

It is also worth noting that in realistic experiments, the initial states in both harmonic traps are
usually identical mixed thermal states as discussed in Ref.~\cite{StevensonPRL2015}, and if the
manipulation of atoms in the phase space during the interrogation is a cyclic evolution of the mixed
state with respect to the center of the probability distribution, then the finite temperature does
not affect both the contrast and the geometric Sagnac phase, where the latter is proportional to the
area difference of the two enclosed trajectories in the phase space. This result shows that the
proposed geometric rotation sensing schemes are not restricted to zero temperature and the initial
single-particle ground state in the harmonic trap. Our work could stimulate further interests and
studies on phase-space geometric quantum sensing with guided matter waves.

\vspace*{-1.ex}
\acknowledgments
\vspace*{-1.5ex}

We thank S. A. Haine for useful communications. This work was supported by
the National Key Research and Development Program of China (No.~2017YFA0304202 and No.~2017YFA0205700),
the NSFC through Grant No.~11475146, and the Fundamental Research Funds for the Central Universities
through Grant No.~2017FZA3005.

\appendix

\begin{widetext}

\section{Derivation of the time-evolution operator $U_{\eta}(t)$ in Eq.~(\ref{eq:U})}
\label{apped:U}

Here we give the detailed derivations of the total time-evolution operator $U(t)$ and the
single-component time-evolution operator $U_{\eta}(t)$ in Eq.~(\ref{eq:U}). With the
properties of projection operators, $\Pi_i \Pi_j =\delta_{ij}\Pi_i$ for $i, j \in \{0, 1\}$,
one can obtain
\begin{eqnarray}
U(T) &=& {\cal{T}} \mathrm{exp}\left[-i\int_0^T H(t)\mathrm{d}t/\hbar\right] \nonumber \\
     &=& {\cal{I}}_{\mathrm{h}} \otimes {\cal{I}}_{\mathrm{s}}+\left(-i/\hbar\right)
     \int_0^T H(t) \mathrm{d}t + \sum_{k=2}^{\infty}\frac{\left(-i/\hbar\right)^k}{k!}
     \int_0^T \mathrm{d}t_0 \int_0^{t_0}\mathrm{d}t_{1}\cdot \cdot \cdot \int_0^{t_{k-2}}
     \mathrm{d}t_{k-1} H\left(t_0\right)H\left(t_1\right)\cdot \cdot \cdot H\left(t_{k-1}\right) \nonumber \\
     &=& {\cal{I}}_{\mathrm{h}} \otimes {\cal{I}}_{\mathrm{s}} + \left \{{\cal{T}}
     \mathrm{exp}\left[-i\int_0^T H_0(t)\mathrm{d}t/\hbar\right]-{\cal{I}}_{\mathrm{h}}\right\} \Pi_0
     + \left \{ {\cal{T}}\mathrm{exp}\left[-i\int_0^T H_1(t)\mathrm{d}t/\hbar\right]
     -{\cal{I}}_{\mathrm{h}}\right \} \Pi_1 \nonumber \\
     &=& U_0(T)\Pi_0+U_1(T)\Pi_1,
\end{eqnarray}
where ${\cal{I}}$ is the identity operator and
$U_{\eta}(T)={\cal{T}} \mathrm{exp}\left[-i\int_0^T H_{\eta}(t)\mathrm{d}t/\hbar\right]$
for $\eta \in \{0, 1\}$, with $H_{\eta}(t)$ being the time-dependent single-component Hamiltonian
for the harmonic oscillator mode, and we have used the relation $\Pi_0 + \Pi_1 = {\cal{I}}_{\mathrm{s}}$.

The Hamiltonian in Eq.~(\ref{eq:Hamiltonian}) describes a forced harmonic oscillator and the
corresponding time-evolution operator at time $t$ can be written as
\begin{equation}
\label{eq:SU}
U_{\eta}(t)= U_{(0)}(t) {\cal{U}}_{\eta}(t),
\end{equation}
where $U_{(0)}(t)=\mathrm{exp}\left[-i\omega_0 \left(a^{\dagger}a +\frac{1}{2}\right)t \right]$ and ${\cal{U}}_{\eta}(t)$
satisfies
\begin{equation}
\label{eq:SDiracU}
i \hbar \frac{\partial}{\partial t} {\cal{U}}_{\eta}(t)
= i\lambda_{\eta}(t)\left[\tilde{a}(t)-\tilde{a}^{\dagger}(t)\right] {\cal{U}}_{\eta}(t),
\end{equation}
where $\tilde{a}(t)=a \mathrm{exp}\left(-i\omega_0 t \right)$. Eq.~(\ref{eq:SDiracU}) can be solved
from the Magnus expansion~\cite{Magnus} and is given by
\begin{equation}
{\cal{U}}_{\eta}(t)= D\left[\beta_{\eta}(t)\right] \mathrm{exp}\left[i\phi_{\eta}(t) \right],
\end{equation}
where $\beta_{\eta} (t) = - \int^t_0 \lambda_{\eta}(\tau) \mathrm{exp}(i\omega_0 \tau)  \mathrm{d} \tau/\hbar$,
\begin{equation}
\label{eq:SPhi}
\phi_{\eta}(t) = \int_0^t \int_0^{\tau_1} \lambda_{\eta}(\tau_1) \lambda_{\eta}(\tau_2) \sin \left[\omega_0
\left(\tau_1 - \tau_2 \right)\right]\mathrm{d}\tau_2 \mathrm{d}\tau_1/\hbar^2,
\end{equation}
and $D\left(\beta\right)=\mathrm{exp}\left(\beta a^{\dagger}-\beta^* a\right)$ is the displacement
operator for the oscillator. Therefore, the time evolution operator in Eq.~(\ref{eq:SU}) reads
\begin{eqnarray}
U_{\eta}(t)&=& U_{(0)}(t) D\left[\beta_{\eta}(t)\right] U_{(0)}^{\dagger}(t)
\mathrm{exp}\left(-i\omega_0 a^{\dagger}a t \right) \mathrm{exp}\left[i\left(\phi_{\eta}(t) - \omega_0 t/2\right)\right] \nonumber \\
           &=&  D\left[\alpha_{\eta}(t)\right] \mathrm{exp}\left(-i\omega_0 a^{\dagger}a t \right)
           \mathrm{exp}\left[i\left(\phi_{\eta}(t) - \omega_0 t/2\right)\right],
\end{eqnarray}
which is Eq.~(\ref{eq:U}) in the main text, with
\begin{equation}
\label{eq:Alpha}
\alpha_{\eta} (t) = \beta_{\eta} (t)  \mathrm{exp}\left(-i\omega_0 t\right)
=- \int^t_0 \lambda_{\eta}(\tau) \mathrm{exp}\left[i\omega_0 \left(\tau-t\right)\right]  \mathrm{d} \tau /\hbar.
\end{equation}

\section{Derivation of the interferometer phase $\phi_I$ in Eq.~(\ref{eq:InterferometerPhase})}
\label{apped:Phase}

In this appendix we present detailed calculations of the interferometer phase $\phi_I$
in Eq.~(\ref{eq:InterferometerPhase}), which establishes a relationship between $\phi_I$
and the well-known Sagnac phase.  The ${\cal{C}}_{1, 0}$ in the spin density matrix
$\rho_{\mathrm{s}}(T)$ is given by
\begin{eqnarray}
\label{eq:SCoeff}
{\cal{C}}_{1, 0} &=& {}_{\mathrm{h}}\hspace{-0.5ex} \left\langle \alpha_1(T) | \alpha_0(T) \right \rangle_{\mathrm{h}}
                      \mathrm{exp}\left\{-i\left[\phi_1(T)-\phi_0(T)\right]\right\} \nonumber \\
                 &=&  \mathrm{exp}\left(-|\Delta \alpha|^2/2\right)
\mathrm{exp}\left\{i\left[\phi_0(T)-\phi_1(T)+\mathrm{Im}\left(\alpha_1^{*}(T)\alpha_0(T)\right)\right]\right\},
\end{eqnarray}
where $\Delta \alpha=\alpha_0 (T)-\alpha_1(T) = -2r \sqrt{\pi m \omega_0/\hbar}
\widetilde{{\cal{W}}}_P^* \left(\omega_0\right)\mathrm{exp}\left(-i\omega_0 T\right)
\propto \widetilde{{\cal{W}}}_P^* \left(\omega_0\right)$.
In general, for an arbitrary time dependent function $\omega_P(t)$ which satisfies
$\int_0^T \omega_P(\tau)\mathrm{d} \tau =\pi$,
the explicit expression for $\phi_{\eta}(T)$ from Eq.~(\ref{eq:SPhi}) is difficult to obtain.
Whereas, in terms of $\widetilde{{\cal{W}}}_P \left(\omega_0\right)$, the phase difference and the
imaginary part in Eq.~(\ref{eq:SCoeff}) have explicit forms, which are given by
\begin{eqnarray}
\label{eq:Sphi01}
\phi_0(T)-\phi_1(T) &=& \int_0^T \int_0^{\tau_1} \left[\lambda_{0}(\tau_1) \lambda_{0}(\tau_2) - \lambda_{1}(\tau_1) \lambda_{1}(\tau_2) \right]
\sin \left(\omega_0 \left(\tau_1 - \tau_2 \right)\right)\mathrm{d}\tau_2 \mathrm{d}\tau_1/\hbar^2     \nonumber \\
                    &=& \left(m \omega_0 \Omega r^2/\hbar\right) \int_0^T \int_0^{\tau_1} \left[\omega_P(\tau_1) +
                    \omega_P(\tau_2) \right]\sin \left(\omega_0 \left(\tau_1 - \tau_2 \right)\right)\mathrm{d}\tau_2 \mathrm{d}\tau_1    \nonumber \\
                    &=& \left(m \Omega r^2/\hbar\right)  \left[2\pi-\left(1+\mathrm{cos} \omega_0 T\right)
                    \int_0^T \omega_P(\tau) \mathrm{cos} \omega_0 \tau \mathrm{d}\tau - \mathrm{sin}
                    \omega_0 T \int_0^T \omega_P(\tau) \mathrm{sin} \omega_0 \tau \mathrm{d}\tau \right]   \nonumber \\
                    &=& \phi_S \left \{ 1-\frac{1}{\sqrt{2\pi}} \mathrm{Re}\left[\left(1+\mathrm{e}^{i\omega_0 T}\right)
                    \widetilde{{\cal{W}}}_P \left(\omega_0\right)\right]\right\},
\end{eqnarray}
and
\begin{eqnarray}
\label{eq:Sarg}
\mathrm{Im}\left(\alpha_1^{*}(T)\alpha_0(T)\right)
&=& \int_0^T \int_0^T \lambda_{0}(\tau_1) \lambda_{1}(\tau_2)
\sin \left(\omega_0 \left(\tau_1 - \tau_2 \right)\right)\mathrm{d}\tau_2 \mathrm{d}\tau_1/\hbar^2     \nonumber \\
&=& \left[m \omega_0 \Omega r^2/\left(2\hbar\right)\right] \int_0^T \int_0^T \left[\omega_P(\tau_1) -
\omega_P(\tau_2) \right]\sin \left(\omega_0 \left(\tau_1 - \tau_2 \right)\right)\mathrm{d}\tau_2 \mathrm{d}\tau_1    \nonumber \\
&=& -\left(m \Omega r^2/\hbar\right)  \left[\left(1-\mathrm{cos} \omega_0 T\right)\int_0^T \omega_P(\tau) \mathrm{cos}
\omega_0 \tau \mathrm{d}\tau - \mathrm{sin} \omega_0 T \int_0^T \omega_P(\tau) \mathrm{sin} \omega_0 \tau \mathrm{d}\tau \right]   \nonumber \\
&=& \frac{\phi_S}{\sqrt{2\pi}} \mathrm{Re}\left[\left(\mathrm{e}^{i\omega_0 T}-1\right)
\widetilde{{\cal{W}}}_P \left(\omega_0\right)\right],
\end{eqnarray}
respectively, where $\phi_S=2m \pi r^2\Omega/\hbar$ is the Sagnac phase and we have used the relation
\begin{eqnarray}
\label{eq:IntMethod}
\int_0^T \int_0^{\tau_1} \omega_P(\tau_2) \sin \left[\omega_0 \left(\tau_1 - \tau_2 \right)\right]\mathrm{d}\tau_2 \mathrm{d}\tau_1
&=& \int_0^T \int_{\tau_2}^T \omega_P(\tau_2) \sin \left[\omega_0 \left(\tau_1 - \tau_2 \right)\right]\mathrm{d}\tau_1 \mathrm{d}\tau_2   \nonumber \\
&=& \int_0^T \int_{\tau_1}^T \omega_P(\tau_1) \sin \left[\omega_0 \left(\tau_2 - \tau_1 \right)\right]\mathrm{d}\tau_2 \mathrm{d}\tau_1.
\end{eqnarray}

Finally, the population difference,
$_{\mathrm{s}} \hspace{-0.3ex} \left\langle \sigma_z \right \rangle _{\mathrm{s}}$,
is given by
\begin{eqnarray}
\label{eq:SSignal}
\quad_{\mathrm{s}} \hspace{-0.3ex} \left\langle \sigma_z \right \rangle _{\mathrm{s}}
&=& \mathrm{Tr}_{\mathrm{s}}\left[\rho_{\mathrm{s}}(T) \sigma_z\right]  \nonumber \\
                         &=& -|{\cal{C}}_{1, 0}|\mathrm{cos}\left(\phi_I\right),
\end{eqnarray}
where the modulus $|{\cal{C}}_{1, 0}|=\mathrm{exp}\left(-|\Delta \alpha|^2/2\right)$ gives the
signal contrast, and the interferometer phase is given by Eq.~(\ref{eq:SCoeff}) and reads
\begin{eqnarray}
\label{eq:SInterferometerPhase}
\phi_I &=& \mathrm{arg} \left( {\cal{C}}_{1, 0} \right) \nonumber \\
       &=&  \phi_0(T)-\phi_1(T)+\mathrm{Im}\left(\alpha_1^{*}(T)\alpha_0(T)\right) \nonumber \\
       &=& \phi_S \left\{1-\sqrt{\frac{2}{\pi}} \mathrm{Re}\left[\widetilde{{\cal{W}}}_P
       \left(\omega_0\right)\right]\right\}.
\end{eqnarray}
With the properties $\omega_P(t) \ge 0$ for $t \in [0, T]$, $\left| \cos \omega_0t\right| \le 1$
and $\int_0^T \omega_P(t) \mathrm{d}t = \pi$, one can obtain
$\left| \mathrm{Re}\left[\widetilde{{\cal{W}}}_P \left(\omega_0\right)\right] \right| \le \sqrt{\pi/2}$.
Therefore, we have $0 \le \phi_I \le 2\phi_S$.

\section{Geometric and dynamic decomposition of the interferometer phase}
\label{apped:PhaseDecomposition}

Here we provide detailed calculations of the geometric and dynamic phase-difference
components in $\phi_I$, which is Eq.~(\ref{eq:PhaseComponents}). The dynamic and geometric phases
$\gamma^{\mathrm{d}}_{\eta}(T)$ and $\gamma^{\mathrm{g}}_{\eta}(T)$ in each trap are given by
\begin{eqnarray}
\label{eq:SDPhase}
 \gamma^{\mathrm{d}}_{\eta}(T)
 &=&-\int_0^T \left\langle \psi_{\eta}(t) |H_{\eta}(t)| \psi_{\eta}(t) \right \rangle \mathrm{d}t  \nonumber \\
 &=& -\int_0^T  \left[\omega_0\left(|\alpha_{\eta}(t)|^2+\frac{1}{2}\right)-\frac{2 \lambda_{\eta}(t)}{\hbar}
 \mathrm{Im} \alpha_{\eta}(t) \right] \mathrm{d}t \nonumber \\
 &=& 2 \phi_{\eta}(T)-\omega_0 \int_0^T  |\alpha_{\eta}(t)|^2 \mathrm{d}t - \frac{1}{2}\omega_0 T,
\end{eqnarray}
and
\begin{eqnarray}
\label{eq:SGPhase}
 \gamma^{\mathrm{g}}_{\eta}(T)
 &=& \frac{i}{2}\int_{\Gamma_{\eta}} \alpha_{\eta}^*  \mathrm{d} \alpha_{\eta} - \alpha_{\eta}
 \mathrm{d} \alpha^*_{\eta} - \mathrm{arg}\left[\left\langle \alpha_{\eta}(T) |{\cal{G}} \right \rangle\right] \nonumber \\
 &=& -\int_0^T \mathrm{Im} \left[ \alpha_{\eta}^*(t) \partial_t \alpha_{\eta}(t)\right] \mathrm{d}t   \nonumber \\
 &=& -\phi_{\eta}(T)+\omega_0 \int_0^T |\alpha_{\eta}(t)|^2  \mathrm{d}t,
\end{eqnarray}
respectively, where $\phi_{\eta}(T)$ is given by Eq.~(\ref{eq:SPhi}) and satisfies
$\phi_{\eta}(T)-\omega_0 T/2 =\gamma^{\mathrm{d}}_{\eta}(T)+\gamma^{\mathrm{g}}_{\eta}(T)$,
and $\mathrm{arg}\left[\left\langle \alpha_{\eta}(T) |{\cal{G}} \right \rangle\right]=0$.

In general, the calculations of explicit expressions for dynamic and geometric phases in each trap
are difficult for an arbitrary $\lambda_{\eta}(t)$. Whereas, the dynamic and geometric phase
differences $\Delta \gamma^{\mathrm{d}}$ and $\Delta \tilde{\gamma}^{\mathrm{g}}$ in Eq.~(\ref{eq:PhaseDecomp})
can be expressed in terms of $\widetilde{{\cal{W}}}_P \left(\omega\right)$ and its derivative at
the trap frequency $\omega_0$, which will be shown below.

With $\omega_0 \int_0^T |\alpha_{\eta}(t)|^2  \mathrm{d}t = \left(\omega_0/\hbar^2\right)
\int_0^T \mathrm{d}t \int_0^t \mathrm{d}\tau_1 \int_0^t \mathrm{d}\tau_2 \lambda_{\eta}(\tau_1)
\lambda_{\eta}(\tau_2) \cos \omega_0 \left(\tau_1-\tau_2\right)$,
and by defining
$ \int_0^T \Delta |\alpha(t)|^2 \mathrm{d}t =\int_0^T \left(|\alpha_{0}(t)|^2 - |\alpha_{1}(t)|^2\right) \mathrm{d}t $,
one can easily obtain
\begin{eqnarray}
 \omega_0\int_0^T  \Delta |\alpha(t)|^2 \mathrm{d}t
 &=& \left(\omega_0/\hbar^2\right) \int_0^T \mathrm{d}t  \int_0^t \mathrm{d}\tau_1 \int_0^t
 \mathrm{d}\tau_2 \left[\lambda_{0}(\tau_1)\lambda_{0}(\tau_2) - \lambda_{1}(\tau_1)\lambda_{1}(\tau_2) \right]
 \cos \omega_0 \left(\tau_1-\tau_2\right)     \nonumber \\
 &=& \left(m \omega_0^2 \Omega r^2/\hbar\right) \int_0^T \mathrm{d}t  \int_0^t \mathrm{d}\tau_1 \int_0^t \mathrm{d}\tau_2
 \left[\omega_P(\tau_1) + \omega_P(\tau_2) \right]\cos \omega_0 \left(\tau_1 - \tau_2 \right)   \nonumber \\
 &=& \phi_S \left\{1-\sqrt{\frac{2}{\pi}} \mathrm{Re}\left[\mathrm{e}^{i\omega_0 T}\widetilde{{\cal{W}}}_P
 \left(\omega_0\right)\right] - \sqrt{\frac{2}{\pi}} \omega_0 T \mathrm{Im}\left[\widetilde{{\cal{W}}}_P
 \left(\omega_0\right)\right]-\frac{\omega_0}{\pi}\int_0^T \tau \omega_P(\tau) \sin \omega_0 \tau \mathrm{d}\tau \right \}  \nonumber \\
 &=& \phi_S \left\{1-\sqrt{\frac{2}{\pi}} \mathrm{Re}\left[\mathrm{e}^{i\omega_0 T}\widetilde{{\cal{W}}}_P
 \left(\omega_0\right)\right]- \sqrt{\frac{2}{\pi}} \omega_0 T
 \mathrm{Im}\left[\widetilde{{\cal{W}}}_P\left(\omega_0\right)\right]+\sqrt{\frac{2}{\pi}}
 \omega_0 \partial_{\omega}\mathrm{Re}\left[ \widetilde{{\cal{W}}}_P\left(\omega\right)\right]_{\omega=\omega_0} \right \},  \nonumber \\
\end{eqnarray}
where we have used the same integration method as in Eq.~(\ref{eq:IntMethod}) to obtain the third
equation and we also have used the relation
$\partial_{\omega}\mathrm{Re}\left[ \widetilde{{\cal{W}}}_P\left(\omega\right)\right]_{\omega=\omega_0}
= -\int_0^T \tau \omega_P(\tau) \sin \omega_0 \tau \mathrm{d}\tau/\sqrt{2\pi}$.
Together with Eqs.~(\ref{eq:Sphi01}), (\ref{eq:Sarg}), (\ref{eq:SDPhase}), and (\ref{eq:SGPhase}),
we obtain
\begin{eqnarray}
\label{eq:SDeltaGphase}
\Delta \tilde{\gamma}^{\mathrm{g}}=\sqrt{\frac{2}{\pi}} \phi_S \omega_0 \left\{\partial_{\omega}
\mathrm{Re}\left[ \widetilde{{\cal{W}}}_P\left(\omega\right)\right]_{\omega=\omega_0}
- T \mathrm{Im}\left[\widetilde{{\cal{W}}}_P\left(\omega_0\right)\right]\right\},
\end{eqnarray}
and
\begin{eqnarray}
\label{eq:SDeltaDphase}
\Delta \gamma^{\mathrm{d}}=\phi_S \left\{1-\sqrt{\frac{2}{\pi}}
\mathrm{Re}\left[\widetilde{{\cal{W}}}_P \left(\omega_0\right)\right]\right\}
- \Delta \tilde{\gamma}^{\mathrm{g}},
\end{eqnarray}
respectively.

\section{Phase-space geometric Sagnac phase---Examples}
\label{apped:GeoSagncExamples}

Here we present several examples for the geometric Sagnac phases with designed $\omega_P(t)$
and the interrogation time $T$, with corresponding Fourier transform analyses.

\emph{Example (i)}: Unconventional geometric Sagnac phase. $(1)$ A sinusoidal angular velocity
$\omega_P(t)=\pi^2 |\mathrm{sin}\left(2\pi t/T\right)|/\left(2T\right)$ with $t \in [0, T]$
gives the Fourier transform
\begin{eqnarray}
\mathrm{Re}\left[\widetilde{{\cal{W}}}_P\left(\omega\right) \right] =
\frac{\sqrt{\pi/2} \cos^2\left(\frac{\omega T}{4}\right) \cos\left(\frac{\omega T}{2}\right)}
{1-\left(\frac{\omega T}{2\pi}\right)^2}, \ \ \ \mathrm{Im}\left[\widetilde{{\cal{W}}}_P\left(\omega\right) \right]
= \frac{-\sqrt{2\pi} \cos^3\left(\frac{\omega T}{4}\right) \sin\left(\frac{\omega T}{4}\right)}
{1-\left(\frac{\omega T}{2\pi}\right)^2}.
\end{eqnarray}
So the condition $\phi_I=\phi_S$ requires that $\omega_0 T = (2L+1) \pi$ or $2(2L+1) \pi$ with $L=0, 1, 2,\cdots$,
and $\widetilde{{\cal{W}}}_P^*\left(\omega_0\right)=0$ requires that $\omega_0 T = 2(2L+1) \pi$
($L=0, 1, 2,\cdots.$). The intersection is $\omega_0 T = 2(2L+1) \pi$
($L=0, 1, 2,\cdots.$). Further calculations show that only the $L=0$ case with $T=2\pi/\omega_0$
can give a solution of $\kappa$ in Eq.~(\ref{eq:DGRatio}), which is $\kappa=8/\pi^2$. Therefore,
the Sagnac phase $\phi_S=8 \Delta \tilde{\gamma}^{\mathrm{g}}/\pi^2$ is an unconventional geometric
phase, by which we mean that the geometric $\phi_S$ also involves a dynamic component~\cite{ZhuPRL2003}.
For the other cases with
$L \ne 0$, $\partial_{\omega} \mathrm{Re}\left[\widetilde{{\cal{W}}}_P\left(\omega\right)\right]_{\omega=\omega_0} \equiv 0$,
and $\phi_S$ is completely dynamic.

$(2)$ A cosinusoidal angular velocity~\cite{HainePRL2016}
$\omega_P(t)=\left(\pi/T\right)\left[1-\mathrm{cos}\left(2\pi t/T\right)\right]$
gives the Fourier transform
\begin{eqnarray}
\mathrm{Re}\left[\widetilde{{\cal{W}}}_P\left(\omega\right) \right] =  \frac{\sqrt{\pi/2}
\sin\left(\omega T\right)}{\omega T\left[1-\left(\frac{\omega T}{2\pi}\right)^2\right]},
\ \ \ \mathrm{Im}\left[\widetilde{{\cal{W}}}_P\left(\omega\right) \right] =  \frac{\sqrt{\pi/2}
\left[\cos\left(\omega T\right)-1\right]}{\omega T\left[1-\left(\frac{\omega T}{2\pi}\right)^2\right]}.
\end{eqnarray}
So the condition $\widetilde{{\cal{W}}}_P^*\left(\omega_0\right)=0$ requires that $\omega_0 T = 2M \pi$
($M=2, 3, 4,\cdots.$), and the corresponding $\kappa$ is given by $\kappa=1-M^2$. Therefore, the Sagnac
phase $\phi_S=\left(1-M^2\right)\Delta \tilde{\gamma}^{\mathrm{g}}$ is also an unconventional geometric
phase.

\emph{Example (ii)}: Pure geometric Sagnac phase with a flat temporal profile for $\omega_P(t)$.
A constant angular velocity $\omega_P(t)=\pi/T$ with $t \in [0, T]$ gives the Fourier transform
\begin{eqnarray}
\mathrm{Re}\left[\widetilde{{\cal{W}}}_P\left(\omega\right) \right] = \sqrt{\frac{\pi}{2}}
\frac{\sin \omega T}{\omega T}, \ \ \ \mathrm{Im}\left[\widetilde{{\cal{W}}}_P\left(\omega\right) \right]
= \sqrt{\frac{\pi}{2}} \frac{\cos \omega T -1}{\omega T}.
\end{eqnarray}
Therefore, $\phi_I=\phi_S$ requires that $\omega_0 T = K \pi$ and the maximization of contrast, i.e.,
$\widetilde{{\cal{W}}}_P^*\left(\omega_0\right)=0$, requires
that $\omega_0 T = 2K \pi$, with $K$ being a positive integer. If the interrogation time is selected
to be $T=2K \pi/\omega_0$ ($K=1, 2, 3\cdots.$), then both of the two requirements are met. For this
case, the solution for $\kappa$ in Eq.~(\ref{eq:DGRatio}) is $\kappa=1$. Furthermore, in this example
$\gamma_{\eta}^{\mathrm{d}}(T)=-K \pi$ for both branches with $\eta=0$ and $1$, which comes from
the zero-energy contribution. So the Sagnac phase in this case only has a purely geometric component.

\end{widetext}


%


\begin{thebibliography}{35}
\bibitem [{Dou()}]{DoubleAreaNotes}%
  \BibitemOpen
  \href@noop {} {}\bibinfo {note} {In the $\alpha$ complex plane, the
  differential vector area subtended by the path is
  $\mathrm{d}\mathbf{A}=\left(1/2\right) \left(\mathrm{Re} \alpha, \mathrm{Im}
  \alpha\right) \times \mathrm{d}\left(\mathrm{Re} \alpha, \mathrm{Im}
  \alpha\right)$, where $\times$ is the vector product. This formula and its
  reexpression with $\alpha$ and $\alpha^*$ can expain the prefactor
  "$-2$".}\BibitemShut {Stop}%
\end{thebibliography}
\end{document}